\documentclass[prl,twocolumn]{revtex4}
\begin{document}

\title{
Global minimum determination of the Born-Oppenheimer surface within density functional theory}

\author{Stefan Goedecker, Waldemar Hellmann}
\affiliation{ Institut f\"{u}r Physik, Universit\"{a}t Basel, Klingelbergstr.82, 4056 Basel, Switzerland \\}
\author{Thomas Lenosky}
\affiliation{ Physics Department, Ohio State University \\ 
1040 Physics Research Building, 191 West Woodruff Avenue, Columbus, Ohio 43210-1117}

\begin{abstract}
We present a novel method, which we shall refer to as the dual minima hopping method (DMHM), that allows us to find 
the global minimum of the potential energy surface (PES) within density functional theory (DFT) for 
systems where a fast but less accurate calculation of the PES is possible. 
This method can rapidly find the ground state configuration of clusters and other complex systems 
with present day 
computer power by performing a systematic search. We apply 
the new method to silicon clusters. Even though these systems have already been extensively 
studied by other methods, we find new configurations for Si$_{16}$, Si$_{18}$ and Si$_{19}$ that 
are lower in energy than those found previously.

\end{abstract}


\maketitle

Determining the structure of a molecule, cluster or crystal is 
one of the most fundamental and important tasks in solid state physics 
and chemistry. Practically all physical properties of a system depend 
on its structure. The structural configurations of a system are determined by the 
Born-Oppenheimer PES, which gives the energy 
of a system as a function of its atomic coordinates. Minima of the PES give stable 
configurations. The global minimum  gives the ground state 
configuration. At low enough temperature the system will be found in this global 
minimum structure assuming that this structure is kinetically accessible. 
Since the zero point energy of different structures varies negligibly,
the determination 
of the ground state structure is equivalent to the mathematical problem of finding the 
global minimum of the PES. 

It is well established that the PES of a condensed matter system can 
be calculated with good accuracy within DFT. 
Nevertheless, DFT methods have not been used up to now
as a standard tool in algorithms that attempt  to
determine the ground state of complex systems because most algorithms for the 
determination of the global minimum require an enormous number of evaluations of the PES. 
Since each evaluation requires a full electronic structure calculation, 
these algorithms are computationally too demanding within the full DFT framework.
A systematic search for the global minimum is however possible 
with cheaper methods such as tight binding and force field methods. 

In summary, with present methods one has either the choice of using methods with a limited 
power of predictability or of doing a constrained search for the global minimum. 
In a  constrained search 
one fixes some atomic positions or imposes some structural motifs, but experience shows that 
the global minimum is often missed in this way. 
To overcome this dilemma several researchers have adopted an approach 
where one first effectuates a systematic search with a method that allows for a fast but 
inaccurate calculation of the PES to obtain some candidate structures. 
Which of the candidate structures is lowest in energy is determined in a second step by 
DFT calculations. As we shall show later this approach is generally not applicable.

Other researches have combined systematic search algorithms 
with DFT methods, but their algorithms required too many DFT calculations 
to be computationally feasible if one wants to find the global minimum. 
R\"{o}thlisberger \textit{et al.}~\cite{rothl} have used simulated 
annealing within DFT to find structural motifs of the mid-size clusters but 
their final lowest energy geometries were obtained by other means.
Yoo and Zeng~\cite{zg27_39}, ~\cite{z_last} have combined 
basin-hopping (BH) with DFT and were able to find new low-lying 
minima for some clusters, among them Si$_{16}$, Si$_{17}$ and Si$_{18}$. 
For Si$_{16}$, they have found a new global minimum structure by 
performing a systematic BH search within DFT. As we shall see, 
both the systematic BH for Si$_{16}$ as well as the constrained BH 
for Si$_{17}$ and Si$_{18}$ within DFT have missed the global minimum.

In this paper we shall present a method that allows for a systematic search 
for the global minimum of the PES of a complex system within 
DFT. The method is a modification of the minima hopping 
method (MHM)~\cite{minhop}.  In the MHM one visits a series of 
local minima until the global minimum is found. The algorithm has a double loop 
structure. In the inner loop one attempts to escape from the current  minimum, 
in the outer loop one accepts or rejects new minima found by successful 
escape attempts. A \textit{history list} keeps track of all minima found. 
A feedback mechanism uses information from this history list to make more vigorous 
escape attempts when the algorithm is revisiting previously found minima thereby preventing
the algorithm from getting trapped in an incorrect minimum.  
The inner escape loop contains two basic steps. The first does a certain number of 
molecular dynamics (MD) moves until one has overcome at least one energy barrier. The second
step consists in performing a standard geometry relaxation to reach the closest minimum
with an accurate method. 

In the ordinary version of the MHM~\cite{minhop} the forces for the MD and for the geometry optimization part are done with the same method. Fast methods 
such as force field or tight binding methods have to be used to limit the computing 
time to an acceptable length. In the modified MHM presented in this paper two different methods 
are combined: a slow but accurate method and a fast but less accurate 
method. The fast method is used for the MD part and for the first 
few steps of the geometry  optimization. The accurate method is then used for the final 
geometry optimization and the evaluation of the energy of the relaxed structure. 
In this way the search for the global minimum is reduced to a relatively small 
number of geometry optimizations with the accurate and expensive method plus a 
much larger number of force evaluations with the fast method. 
Henceforth, we shall refer to this modified minima hopping algorithm, 
that combines the two methods for the calculation of the forces, as the dual minima hopping 
method (DMHM). 

The fact that the input configuration 
for the geometry optimization with the accurate method is a configuration that was 
prerelaxed with the fast method is important for the stability of the entire 
algorithm if the accurate method is a DFT method. DFT 
programs do typically not converge if the input configuration is far from 
any physically reasonable configuration. The prerelaxation with the fast method 
excludes the possibility that a physically unreasonable state is used as an input 
configuration. 
From the previous considerations it might seem advantageous to do a full prerelaxation, 
i.e. to use a minimum of the fast method as the input for the geometry 
optimization with the accurate method.
If the fast method is a reasonable approximation then a local minimum found by it 
will often be close to a local minimum of the accurate method. 

Unfortunately, in general there is no one-to-one correspondence between 
minima obtained from the two methods.  Therefore, some minima obtained using the 
accurate method are inaccessible from the starting configurations provided by the fast
method.  
For this reason only a small number 
of steps should be done in the prerelaxation with the fast method. In this way the 
ensemble of the starting configurations for the geometry optimization with the 
accurate method comprises a considerable part of the configurational space (and not only 
the ensemble of all the minima of the fast method) and one can reach virtually any minimum of 
the accurate method.

\begin{figure}[h]             
\begin{center}
\setlength{\unitlength}{1cm}
\begin{picture}( 11.2,5.0)           
\put(-1.4,-1.2){\includegraphics{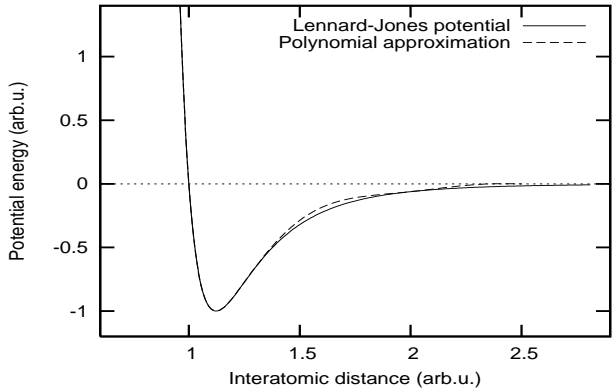}}   
\end{picture}
\caption{ \label{ljpots} The truncated polynomial approximation of the Lennard-Jones potential and the exact Lennard-Jones potential. }
\end{center}
\end{figure}

The Bell-Evans-Polanyi (BEP) principle ~\cite{F_Jensen} states that highly exothermic chemical
reactions have a low activation energy. In the context of a global minimum search, 
this means energetically low configurations will be preferably 
found behind low energy barriers.
The BEP principle is essential for the success of the MHM as has been shown in ~\cite{minhop}. 
The correlation between the barrier height and the energy 
of the minimum `behind' the barrier certainly deteriorates if one is combining two 
different methods. 
This implies more local minima will be visited, on average, with the DMHM 
before the global minimum is found than with the ordinary MHM. 
In order to explore 
the influence of 
this reduced correlation we did systematic tests with a 38 atom Lennard-Jones (LJ) cluster.
This is a system for which the global minimum is hard to find 
since it is contained in a small secondary funnel~\cite{wales}, but the 
computing time is small since the potential can be evaluated very rapidly.
As the accurate method we used the LJ potential. As the `fast' method we 
used a truncated polynomial 
approximation of the LJ potential as shown in Fig.~\ref{ljpots}. 
As expected, the number of local minima that are visited on average before the global 
minimum is found increases from 380 to 530, nevertheless, the 
number of force evaluations needed with the `expensive' exact LJ method 
is reduced by a factor of 5. 

\begin{figure}[h]             
\begin{center}
\setlength{\unitlength}{1cm}
\begin{picture}( 20.0,8.5)           

\put(-0.3,5.8){\includegraphics{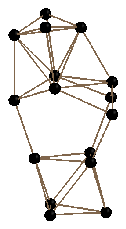}}
\put(1.0,5.9){Si$_{16}$}

\put(-0.5,2.4){\includegraphics{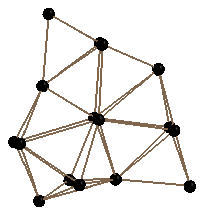}}
\put(1.0,2.7){Si$_{16a}$}

\put(2.4,5.63){\includegraphics{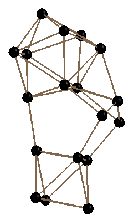}}
\put(3.9,5.9){Si$_{18}$}

\put(2.5,2.35){\includegraphics{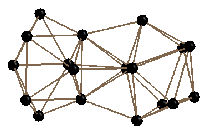}}
\put(4.0,3.0){Si$_{18a}$}

\put(5.2,5.5){\includegraphics{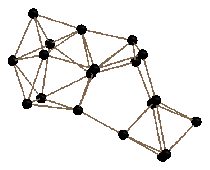}}
\put(6.5,6.3){Si$_{19}$}

\put(5.1,2.55){\includegraphics{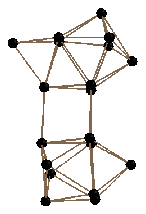}}
\put(6.8,2.9){Si$_{19a}$}

\put(-0.4,-0.6){\includegraphics{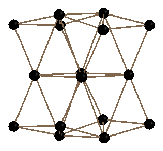}}
\put(1.0,-0.1){Si$_{16b}$}

\put(2.5,-0.37){\includegraphics{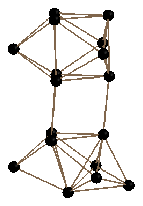}}
\put(4.0,-0.1){Si$_{17a}$}

\put(5.5,-0.3){\includegraphics{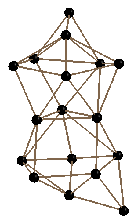}}
\put(6.8,-0.1){Si$_{17b}$}

\end{picture}
\caption{ \label{clusters} Lowest energy geometries Si$_{16a}$, Si$_{16b}$, Si$_{17a}$, Si$_{17b}$, Si$_{18a}$ and 
           Si$_{19a}$ found in this work with DMHM and the putative global minimum structures Si$_{16}$~\cite{z_last},
           Si$_{18}$~\cite{z_12_20},~\cite{Jackson_1} and Si$_{19}$~\cite{Jackson_1} reported previously. The new geometries 
           will be posted on the Cambridge Cluster Database ~\cite{cam_databse}.}
\end{center}
\end{figure}

To demonstrate that the DMHM can indeed find the global ground state geometry 
of real clusters, we have applied it to silicon clusters. Numerous  groups are involved in 
the search of the ground state of silicon clusters and there are at least 50 
theoretical papers on this subject ~\cite{rothl}-~\cite{z_last},~\cite{Ho_Nature}-~\cite{Li_70}.  
Applying DMHM to silicon clusters we were able to find within several days of computing 
time all of the known structures ~\cite{sank},~\cite{fourn},~\cite{z_7_11}, ~\cite{z_12_20}
in the range Si$_{4}$-Si$_{19}$ and we even found 
lower energy structures for Si$_{16}$, Si$_{18}$ and
Si$_{19}$ in spite of the fact that silicon clusters up to 19 atoms in size have already been 
extensively studied. The new global minimum structures within CPMD/PBE (see below) Si$_{16a}$, Si$_{18a}$ and 
Si$_{19a}$ as well as the new low-lying isomers Si$_{16b}$, Si$_{17a}$ and Si$_{17b}$ are shown 
in Fig.~\ref{clusters}. 
The structure Si$_{16a}$ contains the TTP-Si$_{9}$-subunit~\cite{TTP_1} and is compact in contrary 
to the structure Si$_{16}$ reported by X.C.Zeng~\cite{z_last}. The structure Si$_{18a}$ is prolate and 
consists of two TTP-Si$_{9}$-subunits which are rotated against each other. The structure Si$_{19a}$ 
consists of a TTP-Si$_{9}$-subunit and a Si$_{10}$-subunit. The low-lying isomer Si$_{16b}$ is compact 
and highly symmetric. The low-lying isomer Si$_{17a}$ consists of a TTP-Si$_{9}$-subunit and a Si$_{8}$-subunit. 
The low-lying isomer Si$_{17b}$ consists of two equal 7-blocks, which are rotated 
against each other, and a triangle as a cleaving block. In contrast to the previous works, 
our configurations were found by the DMHM automatically after having visited only a few hundred local minima. 

As the fast method we have used the Lenosky tight binding scheme for silicon~\cite{Len_TB}. The accurate 
method is DFT as implemented in the Quickstep code~\cite{quickstep}. 
After having performed the DMHM with Quickstep using a relatively small Gaussian basis set and the local 
density approximation (LDA), we have calculated accurate final energies and 
zero-point energies with the CPMD program~\cite{cpmd} using the PBE functional~\cite{pbe}, a high accuracy 
pseudo-potential~\cite{mypsp}, large super-cells (24 \AA) and a sufficient plane wave cutoff (28 Rydberg).
The results for Si$_{16a}$, Si$_{18a}$ and Si$_{19a}$ as compared to Si$_{16}$, Si$_{18}$ and Si$_{19}$ 
are presented in Table~\ref{eclusters}. The low-lying isomer Si$_{16b}$ is virtually isoenergetic with the 
structure Si$_{16}$. The isomer Si$_{17a}$ is lower by 0.16 eV, the isomer Si$_{17b}$ is lower by 0.06 eV 
within CPMD/PBE than the new low-lying Si$_{17}$ isomer reported by X.C.Zeng ~\cite{z_last}. The Si$_{17}$ structure 
reported by Ho \textit{et al.} in ~\cite{Ho_Nature} is however 0.10 eV lower than our isomer Si$_{17a}$. 
In contrast to other exchange correlation functionals, the PBE functional ~\cite{pbe} was not fitted to any
chemical systems with simple bond structures and is expected to give the most accurate description of
the complex bonding patterns found in silicon clusters. The term `accurate' must be handled with caution however, 
since DFT is only an approximation and, as a matter of fact, the energetic ordering may change if one uses 
different functionals ~\cite{z_12_20}.
   
\begin{figure}[h]             
\begin{center}
\setlength{\unitlength}{1cm}
\begin{picture}( 11.0,7.0)           
\put(-0.75,-1.0){\includegraphics{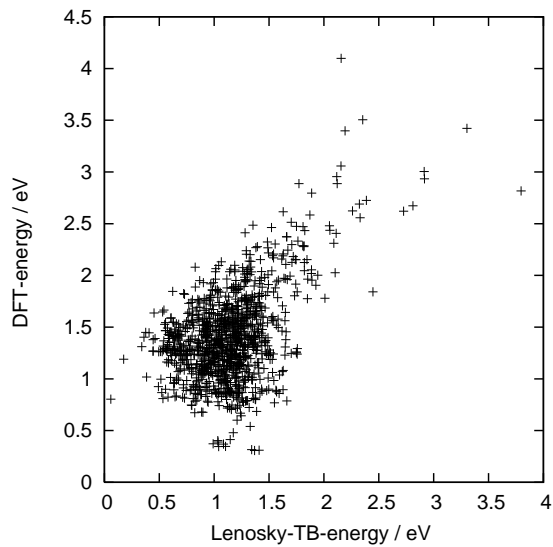}}   
\end{picture}
\caption{ \label{tbdft} The correlation between tight binding and density functional
energies for various configurations of a Si$_{25}$ cluster. If the correlation
was perfect
all the points would lie on the diagonal. Instead the scattering shows
that the tight binding energies can predict the energy differences
between various cluster configurations only with an error of about
1 eV.}
\end{center}
\end{figure}

\begin{table}
\begin{tabular}{| p{1.8cm} | p{1.2cm} | p{1.2cm} | p{1.2cm} |} \hline
\hspace*{0.2cm} Cluster &  \hspace*{0.26cm} Si16 &  \hspace{0.26cm} Si18 &   \hspace{0.26cm} Si19  \\ \hline
\hspace*{0.2cm} PBE     &  \hspace*{0.26cm}-0.15 &  \hspace{0.26cm}-0.01 &   \hspace{0.26cm}-0.08  \\ \hline
\hspace*{0.2cm} PBE(Z)  &  \hspace*{0.26cm}-0.16 &  \hspace{0.26cm}-0.07 &   \hspace{0.26cm}-0.09  \\ \hline
\end{tabular} \\
\caption{ \label{eclusters} The energy differences in eV without and with zero-point energy correction between the 
           lowest energy geometries Si$_{16a}$, Si$_{18a}$  and Si$_{19a}$ found in this work with DMHM and the 
           putative global minimum structures Si$_{16}$~\cite{z_last}, Si$_{18}$~\cite{z_12_20},~\cite{Jackson_1} 
           and Si$_{19}$~\cite{Jackson_1} reported previously using the PBE exchange-correlation 
           functional as implemented in CPMD.}
\end{table} 
Among the various force fields and tight binding schemes we have tested, the Lenosky 
tight binding scheme~\cite{Len_TB} gave the best agreement with the DFT energies. 
It can predict the DFT energies with an error of roughly 
1 eV as shown in Fig.~\ref{tbdft}. Fig.~\ref{tbdft} also shows why the common 
approach of first finding candidate structures by doing a systematic search with a cheap 
method and then checking by an accurate method which of the candidate structures gives 
the global minimum is problematic except for very small systems.  
For a 25 atom silicon cluster the number of geometric configurations within
1 eV above the ground state is of the order of $10^4$ states, for a 33 atom
cluster it is already of the order of $10^5$ states and it 
increases exponentially with system size. 
It is therefore virtually impossible to check which out of these
$10^4$  to $10^5$ configurations is the global minimum in
DFT. Besides, because of the absence of the one-to-one correspondence  
between the local minima of the fast method and of the accurate 
method, it is not guaranteed that any of the minima of the fast method will lead 
to the global minimum of the accurate method upon relaxation.

The identification of the previously visited minima is an essential ingredient of the MHM. 
In the context of the ordinary MHM the energy can be used 
to identify configurations since it is possible to calculate the energy with many significant 
digits both for force fields and tight binding schemes. With DFT programs 
this is not any more possible because of the presence of numerical noise. For this reason we 
have used in addition to the energy all inter-atomic distances. Two DFT minima are considered 
to be identical if all their inter-atomic distances ordered by magnitude agree to within a certain 
tolerance. 

In summary, we have presented a method that allows one to find the global minimum of the 
DFT potential energy surface within acceptable computer time for 
moderately complex systems. The method is efficient for the following reasons.
First, it requires only DFT calculations for configurations where DFT programs typically 
converge without problems. It does not, for instance, require DFT calculations for 
configurations generated by random displacements from a previous configuration.
Second, the MHM is highly efficient in the sense that the number of minima visited 
before the ground state is found is small. Even though the DMHM is not quite as good 
from this perspective it is still efficient if the fast method used for the MD part 
is qualitatively correct. Third, most of the force evaluations are done with 
the fast method and the total effort for finding the global minimum is equal to the effort 
of doing only an affordable number of geometry optimizations with the accurate method.

We thank the Swiss National Science Foundation for the financial support of our research work.
We are grateful to the staff of the computing center at the university of Basel and especially 
to M. Jacquot for technical support and assistance. We thank X.C. Zeng for providing
us the data on his Si$_{16}$, Si$_{17}$ and Si$_{18}$ clusters and on the Si$_{18}$ cluster 
of I. Rata as well as for valuable comments. We thank K.A. Jackson for providing us his data 
on the Si$_{19}$ cluster. We are grateful to M. Krack for his help with the Quickstep-Code. 
We thank C. Umrigar, C. Bruder and M. Rayson for useful comments on the manuscript. The work 
at Ohio State was supported by DOE-DE-FG02-99ER45795 and NSF-ITR-0326386.

\end{document}